\documentclass[]{aastex631}

\usepackage{multirow,colortbl}

\newcommand{\Vm}{V$_\mathrm{mean}$}
\usepackage{multirow}
\usepackage{rotating}
\received{April 15, 2024}
\revised{May 20, 2024}
\accepted{May 31, 2024}

\submitjournal{ApJ}

\graphicspath{{./}{figures/}}
\begin{document}

\title{Inter-planetary type-IV solar radio bursts: A comprehensive catalog and statistical results}

\correspondingauthor{Atul Mohan}
\email{atul.mohan@nasa.gov}

\author[0000-0002-1571-7931]{Atul Mohan}
\affiliation{NASA Goddard Space Flight Center, Greenbelt, MD 20771, USA}
\affiliation{The Catholic University of America, Washington, DC 20064, USA}

\author[0000-0001-5894-9954]{Nat Gopalswamy}
\affiliation{NASA Goddard Space Flight Center, Greenbelt, MD 20771, USA}

\author[0000-0001-5742-9033]{Anshu Kumari}
\affiliation{NASA Goddard Space Flight Center, Greenbelt, MD 20771, USA}
\author[0000-0002-7281-1166]{Sachiko Akiyama}
\affiliation{NASA Goddard Space Flight Center, Greenbelt, MD 20771, USA}
\affiliation{The Catholic University of America, Washington, DC 20064, USA}

\author[0000-0002-5893-1938]{Sindhuja G}
\affiliation{NASA Goddard Space Flight Center, Greenbelt, MD 20771, USA}
\affiliation{The Catholic University of America, Washington, DC 20064, USA}
\affiliation{Bangalore University, Mysore Rd, Jnana Bharathi, Bengaluru, Karnataka, IN}
%% Note that the \and command from previous versions of AASTeX is now
%% depreciated in this version as it is no longer necessary. AASTeX 
%% automatically takes care of all commas and "and"s between authors names.

%% AASTeX 6.31 has the new \collaboration and \nocollaboration commands to
%% provide the collaboration status of a group of authors. These commands 
%% can be used either before or after the list of corresponding authors. The
%% argument for \collaboration is the collaboration identifier. Authors are
%% encouraged to surround collaboration identifiers with ()s. The 
%% \nocollaboration command takes no argument and exists to indicate that
%% the nearby authors are not part of surrounding collaborations.

%% Mark off the abstract in the ``abstract'' environment. 
\begin{abstract}
Decameter hectometric (DH; 1-14 MHz) type-IV radio bursts are produced by flare-accelerated electrons trapped in post-flare loops or the moving magnetic structures associated with the CMEs.
From a space weather perspective, it is important to systematically compile these bursts, explore their spectro-temporal characteristics, and study the associated CMEs.
We present a comprehensive catalog of DH type-IV bursts observed by the Radio and Plasma Wave Investigation (WAVES) instruments onboard Wind and STEREO spacecraft, covering the period of white-light CME observations by the Large Angle and Spectrometric Coronagraph (LASCO) onboard the SOHO mission between November 1996 and May 2023.
The catalog has 139 bursts, of which 73\% are associated with a fast ($>$\,900\,km\,s$^{-1}$) and wide ($>60^\circ$) CME, with a mean CME speed of 1301\,km\,s$^{-1}$. 
{ All DH type-IV bursts are white-light CME-associated, with 78\% of the events associated with halo CMEs.}
The CME source latitudes are within $\pm$45$\circ$. 
77 events had multi-vantage point observations from different spacecraft, letting us explore the impact of line of sight on the dynamic spectra. { For 48 of the 77 events, there was good data from at least two spacecraft.} 
{ We find that, unless occulted by nearby plasma structures, a type-IV burst is best-viewed when observed within $\pm$60$^\circ$ line of sight.}
Also, the bursts with a duration above 120\,min, have source longitudes within $\pm$60$^\circ$. { Our inferences confirm the inherent directivity in the type-IV emission. Additionally, the catalog forms a sun-as-a-star DH type-IV burst database.}
\end{abstract}

%% Keywords should appear after the \end{abstract} command. 
%% The AAS Journals now uses Unified Astronomy Thesaurus concepts:
%% https://astrothesaurus.org
%% You will be asked to selected these concepts during the submission process
%% but this old "keyword" functionality is maintained in case authors want
%% to include these concepts in their preprints.
\keywords{Active sun(18) --- Solar coronal mass ejections(310) --- Stellar coronal mass ejections(1881) --- Solar radio flares(1342) --- Space weather(2037) --- Catalogs(205)}

%% From the front matter, we move on to the body of the paper.
%% Sections are demarcated by \section and \subsection, respectively.
%% Observe the use of the LaTeX \label
%% command after the \subsection to give a symbolic KEY to the
%% subsection for cross-referencing in a \ref command.
%% You can use LaTeX's \ref and \label commands to keep track of
%% cross-references to sections, equations, tables, and figures.
%% That way, if you change the order of any elements, LaTeX will
%% automatically renumber them.
%%
%% We recommend that authors also use the natbib \citep
%% and \citet commands to identify citations.  The citations are
%% tied to the reference list via symbolic KEYs. The KEY corresponds
%% to the KEY in the \bibitem in the reference list below. 

%Origin of High-Energy Protons Responsible for Late-Phase Pion-Decay Gamma-Ray Continuum from the Sun
\section{Introduction} \label{sec:intro}
Type-IV radio bursts were first classified in the meterwaveband by \cite{1957CRAS..244.1326B}. They appear as broadband continuum features in the dynamic spectrum and come in two flavours: stationary type-IV and moving type-IVs~\citep{takakura61_typeIVtypes,young61_DScharacteriation_solarbursts,kundu63_DScharacterisationTypeIV}. A moving type-IV burst shows a clear drift in the dynamic spectrum while the stationary type-IV bursts do not.
The sources of moving type-IV bursts were soon recognised to be moving away from the sun~\citep{Weiss63_typeIV_plasmaEmiss,Boischot68_Bfield_frmtypeIVchar}.
%These bursts were initially thought to be associated with certain class of solar flares \citep{1963SSRv....2...70F}. 
Using simultaneous white-light and radio (80\,MHz) observations of the K-corona, \cite{hansen71coronaltransient_typeII-IV} demonstrated that moving type-IV bursts are associated with bright coronal transients and derived a speed of 1300\,km\,s$^{-1}$.
Moving type-IV bursts were soon recognised as produced by flare accelerated electrons trapped in moving magnetised plasmoids associated with coronal mass ejections~\citep[e.g.,][]{takakura61_accle_typeiv,kundu63_DScharacterisationTypeIV,1971SoPh...20..438D, smerd71_mtypeIV_movingplasm, schahl72_mtypeIV_model}. 
\cite{robinson78_mtypeIV_stats} performed a statistical analysis of 23 moving type-IV bursts observed by Culgoora radioheliograph and estimated the source speeds to lie within 200 - 1500\,km\,s$^{-1}$.
Meanwhile, the stationary type-IVs are produced by accelerated electrons trapped in post-flare loops~\citep[see,][for an overview]{wild1970,Mclean85_book}.
The emission mechanism of the type-IV bursts are believed to be primarily via a plasma emission process based on observed polarisation and brightness temperature levels~\citep[e.g.,][]{Weiss63_typeIV_plasmaEmiss,Duncan81_mtypeIV_plasmaemiss_mech,gopal89_mtypeIV_plasmaemiss_modelevent}.
%Har16_typIVm_CME_assoc,Melnik18_typeIVBehindLimb,Vasanth19_typIVmov_img, Liu2022}.
Using 15\,years of metric burst data from Culgoora radioheliograph, \cite{cane88_typII_IV_stats} showed that $\sim$88\% of the 227 metric type-IV bursts, { were associated with a type-II burst, which is produced by electrons accelerated by coronal mass ejection (CME) shocks}~\citep[e.g.][]{Smerd70_typeII_curvedpathTypII,wild1970,gopal05_m-kmTypeII-EnergeticLink, kumari2023b}.
However, only around a third of the type-II bursts had a type-IV link.
Type-IVs are closely connected to strong soft X-ray flares (on average with M class flares), especially ones which last for around an hour or longer, and fast CMEs~\citep{Cane88_SXR_metricburst_props}.
Similarly around 81\% of the type-IVs in cycle 24 were found to be associated with white-light CMEs~\citep{Anshu21_C24typIV_CME_corr}. However, some of their cases of non-associations could be influenced by data gaps or stealth CMEs~\citep[e.g.,][]{Robbrecht09_stealthCME,Dhuys14_ObsChar_stealthCME,Morosan2021} which are known to have no white-light signatures. In such a case the percentage of association can increase.

Decameter-hectometric (DH) type-IV bursts are rare continuations of the metric type-IVs into the inter-planetary space, are almost always associated with fast ($>$900\,kms$^{-1}$) and wide ($>$60$^{\circ}$ angular extent) coronal mass ejections (CMEs) and there by to solar energetic particle events (SEPs)~\citep{gopal04_IPbursts_overview,Gopal11_PREconf,Hillaris16_typeIV}.
{ Unlike the metric type-IVs, DH type-IVs are always associated with CMEs that have a mean speed (\Vm) of $\sim$1500\,kms$^{-1}$ and are $\sim$75\% of the time halo CMEs~\citep{Gopal11_PREconf}. 
About 89\% of these bursts are associated with GOES M- and X- class flares~\citep{Hillaris16_typeIV}.
 Exploring the CMEs, between 1996 and 2006, that caused an SEP and a radio burst in the decimeter to kilometer (km) band, \cite{Miteva17_SEP-radburst_link} showed that 
%of all the SEP associated radio bursts in decimeter to kilometer range, 
the median X-ray flare strength and CME speed are the highest for the events associated with DH - km type-IV bursts. }
DH type-IVs are observed below 14\,MHz and occasionally extend { down} to 1\,MHz with a mean extent down to $\sim$7\,MHz~\citep{Gopal11_PREconf}. They last typically on average for an hour, occasionally extending over days~\citep{Hillaris16_typeIV}. 
Since emission at frequencies below $\sim$10\,MHz cannot to be observed from ground due to ionospheric cut-off, these events can be detected only with space-based instruments~\citep{Gopal11_PREconf}. The Radio and Plasma Wave Investigation \citep[WAVES;][]{WAVES1995} instrument on-board Wind spacecraft and the SWAVES instrument on board Solar TErrestrial RElations Observatory \citep[STEREO;][]{SECCHI2008} spacecraft are the only window into this interplanetary radio emission.
Despite the close association of DH type-IV bursts with SEPs, strong X-ray flares and fast and wide CMEs, { there have been only a few} studies of DH type-IV events compared to other type of solar radio bursts. This is primarily because these events are relatively rare~\citep{Hillaris16_typeIV}, making it difficult to obtain a significant sample to explore their statistical behaviour in a robust manner.
The results mentioned above on the DH type-IV bursts had only $\sim$40 events mainly from just one solar cycle (C23) primarily from Wind spacecraft.
Besides, \cite{gopal16_typeIVdirectivity} had reported that the DH type-IV emission was directed within a cone angle of $\pm$60$^{\circ}$ which is possibly associated with the line of sight occultation of the post flare loop by the solar disk. 
Using a few case studies with multi-vantage point data from Wind and STEREO missions \cite{Nasrin18_DHtypeIV_occult_streamerCMEshock} argued that the visibility of the the DH type-IV could be hampered by the presence of high density structures along the line of sight, which are formed by the interaction of the evolving CME with other pre-existing coronal magnetic field structures like streamers. These high density regions can attenuate the DH band radio emission from the post flare loop.
However, to make robust conclusions on various characteristics of DH type-IVs and the properties of the CMEs/flares that correlate with the observed burst features, we need a statistically complete and significant sample. 
%Such a comprehensive catalog of Wind and STEREO detected DH type-IVs and the associated CMEs/flares would cover two full solar cycles (C23 and C24) and the start of the current cycle (C25), hopefully providing a statistically significant sample for exploration. 

In this work, we aim to put together a comprehensive sample of DH type-IV bursts detected by Wind and STEREO spacecraft, and catalog the properties of the associated CMEs. The catalog covers two full solar cycles (C23 and C24) and the start of the current cycle (C25), providing a statistically significant sample for exploration. 
%via a comprehensive search across the dynamic spectra from Wind, STEREO-A and STEREO-B spacecraft during times when either a white-light CME, a metric type-IV or a DH type-II burst was reported since 1996. 
The DH band spectro-temporal properties of these bursts will be explored alongside the properties of their associated CMEs.
We will also address the issue of directivity in the events detected simultaneously by at least two widely-separated spacecraft.
{ Recently, \cite{Patel2021} studied DH type-II radio bursts during the past two solar cycles.
Since type-IV and type-IIs are generally considered as a radio diagnostic of strong CMEs this work will complement the aforementioned study forming a combined unique database to particle acceleration signatures of inter-planetary CMEs.
Besides, type-IV radio bursts are the only solar-type bursts to be reported so far on other active stars~\citep{Zic20_typeIV_ProximaCen,Atul24_ADLeoTypeIV}.
Since our observations are full solar disk integrated emission, this database will form the first comprehensive sun-as-a-star catalog of interplanetary type-IV bursts.}

Section~\ref{sec:data} presents the sources of data while Sect.~\ref{sec:method} describes the steps involved in identifying and compiling the DH type-IV events from the data sources. Section~\ref{sec:results} presents the DH type-IV catalog and the statistical properties of these events. Section~\ref{sec:discussion} analyse the results and discuss their implications. The conclusions are given in Sect.~\ref{sec:conclusions}.

\section{Data} %%%%%%%%%%%%%%
  \label{sec:data}
The data primarily comes from the compilation of the Coordinated Data Aanalysis Workshops (CDAW) Data Center\footnote{\href{https://cdaw.gsfc.nasa.gov/}{https://cdaw.gsfc.nasa.gov/}}. 
This work made use of the radio dynamic spectra from Wind, STEREO-A and STEREO-B spacecraft in the frequency range 0.02 - 14\,MHz.
The properties of the white-light CMEs observed by Large Angle and Spectrometric COronagraph \citep[LASCO;][]{LASCO1995} onboard Solar and Heliospheric Observatory\footnote{\href{https://www.eoportal.org/satellite-missions/soho}{https://www.eoportal.org/satellite-missions/soho}} (SOHO) are obtained from the LASCO CME catalog \citep{Yashiro04_LASCOCME_Catalog,Gopal09_LASCO_CMEcatalog}. We relied on the event reports from Space Weather Prediction Center\footnote{\href{https://www.swpc.noaa.gov/products/solar-and-geophysical-event-reports}{https://www.swpc.noaa.gov/products/solar-and-geophysical-event-reports}} (SWPC) to obtain the list of metric type-IV events since 1996. 
Besides, in some cases, we relied on various other public data sources to identify metric counterparts of DH type-IVs, namely eCallisto\footnote{\href{https://www.e-callisto.org/links.html}{https://www.e-callisto.org/links.html}}, Radio Monitoring\footnote{\href{https://secchirh.obspm.fr/spip.php?rubrique8}{https://secchirh.obspm.fr/spip.php?rubrique8}} and Australian Space Weather Database\footnote{\href{https://www.sws.bom.gov.au/World\_Data\_Centre/1/1}{https://www.sws.bom.gov.au/World\_Data\_Centre/1/1}}.
The DH type-II catalog~\citep{Gopal19_DHtypeII_catalog} compiled by CDAW Data Center is also used, since they are often associated with type-IVs and CMEs.
\section{Methodology}
\label{sec:method}
%A search during times when LASCO CME catalog reported an event.
{ We adopt a strategy of bias-free blind search for DH type-IV burst signatures. Since DH type-IV bursts are very closely associated with CMEs, the initial list of potential date-time periods to search for DH type-IV bursts was provided by the LASCO CME catalog.}
%Thus, the LASCO CME date-times give us the initial list of potential DH type-IV burst periods.
{ However, we may still miss some DH type-IV events due to potential data gaps in the LASCO data stream or due to events being not associated with white-light CMEs.}
%Besides, we also made a note of the periods when the DH type-II bursts and metric type-IVs were reported. 
To cover such periods and ensure sample completeness, we relied on two other catalogs of bursts potentially associated with DH type-IV bursts: metric type-IV event reports and the DH type-II catalog. 
Since DH type-IVs are continuations of metric type-IVs, the date-time periods from SWPC metric type-IV reports should provide a larger set of potential DH type-IV events.
Nevertheless, the reason why we did not solely rely on the SWPC metric type-IV reports to search for DH type-IVs is that there have been cases where radio bursts go unreported in the SWPC event list.
The DH type-II catalog, being an independent DH band catalog of CME associated radio bursts also provide additional date-time intervals helping us cover up for periods of LASCO data gap.
%Due to this issue, we had to often search for metric DS from various other ground based observatories mentioned in Sec.~\ref{sec:data} to detect the metric counterparts.
%Unlike space based data, it is difficult to get a good quality 24\,h monitoring database of metric radio dynamic spectra from 1996 to present due to observing limitations from the ground.
The date-time interval we chose for the study extends from Nov, 1996 to May, 2023, which is when LASCO data started being available. This ensures that we have a well defined, systematically compiled catalog of CME characteristics covering multiple solar cycles~\citep[][]{Yashiro04_LASCOCME_Catalog,Gopal09_LASCO_CMEcatalog}.
The final list of DH type-IVs search periods include date-times corresponding to 34083 LASCO CMEs, 939 metric type-IVs and 140 DH type-IIs.
\begin{figure}[ht!]
\epsscale{0.8}
\plotone{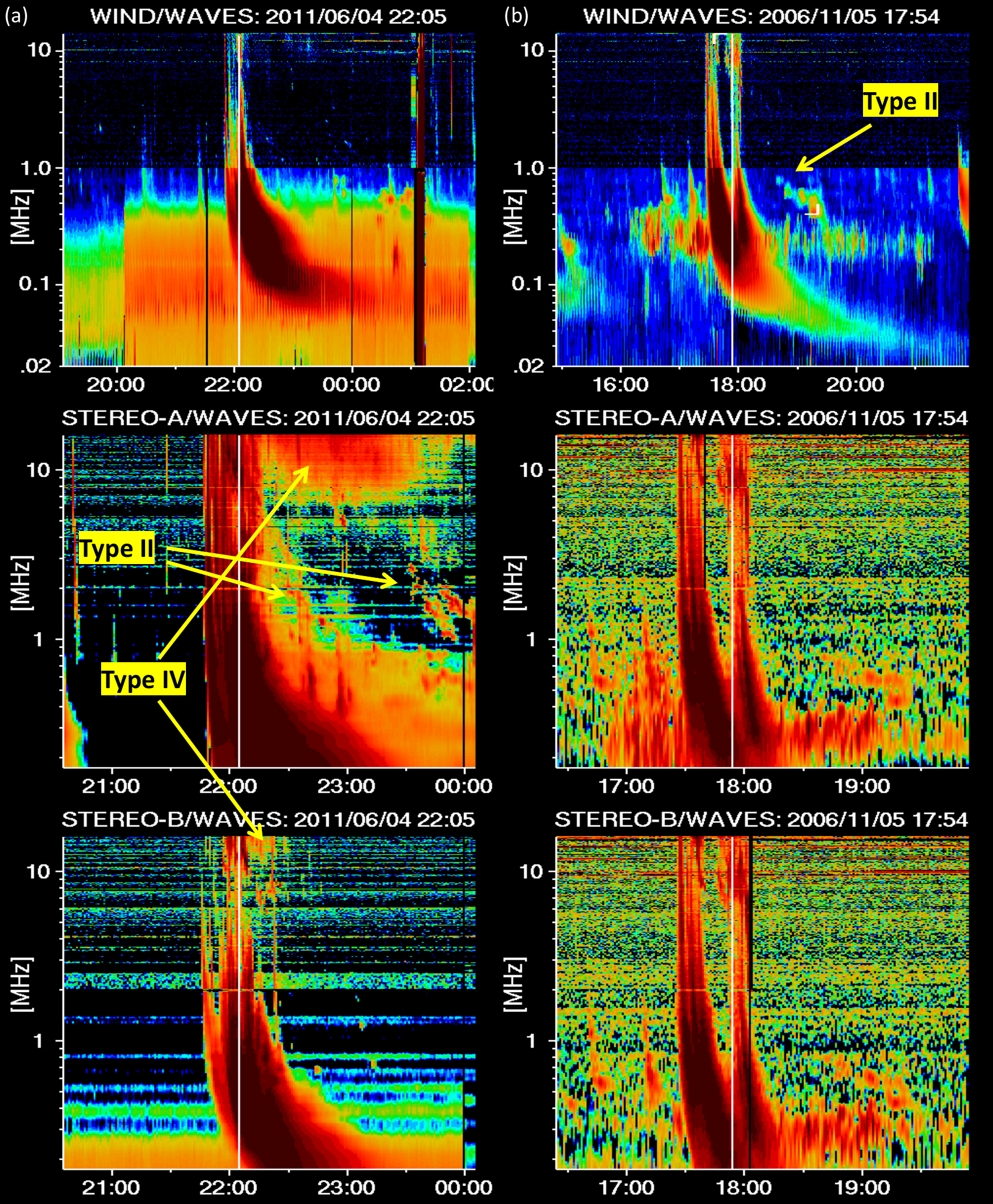}
\caption{Example DH dynamic spectra. (a): A type-IV detection in STEREO-A and B. (b): No type-IV burst, but a type-II is detected by multiple spacecraft. The time periods of the associated LASCO CME events are marked by vertical white lines. 
\label{fig1:example}}
\end{figure}

A Python code was written to gather the dynamic spectra (DS) from Wind and STEREO for each date-time entry in the search list, by querying the respective databases\footnote{\href{https://cdaw.gsfc.nasa.gov/images/wind/}{https://cdaw.gsfc.nasa.gov/images/wind/}}{}$^,$\footnote{\href{https://cdaw.gsfc.nasa.gov/images/stereo/}{https://cdaw.gsfc.nasa.gov/images/stereo/}}. Whenever there is data from multiple missions, a combined DS is made. All the images were manually examined later to identify DH type-IV candidates.
Figure~\ref{fig1:example} shows example DH dynamic spectra from periods with and without a type-IV event.
After identifying the DH-type-IV candidates, the metric dynamic spectra were also accumulated for all events and ensured that there is a potential continuation of the DH type-IV into the metric band.
We removed a couple of very long duration type-IV bursts extending over 10\,h because of the difficulty in associating them with a particular CME event. These events can cause ambiguity in the scientific inferences derived on relationship between DH type-IV characteristics, particularly its duration and properties of associated CMEs.
Thus we ended up with a preliminary sample of DH type-IV events, each of which is well associated with a CME.
%The final list of DH type-IV events contain 139 events, well associated with a CME event.
The preliminary DH type-IV sample was compared against existing event lists used by previous studies~\citep{Gopal11_PREconf,gopal16_typeIVdirectivity,Hillaris16_typeIV} to ensure completeness across them. 
It was also ensured that the metric dynamic spectra, whenever available, from a ground based data repository had signatures of type-IV counterparts.
Some spurious candidates with no clear metric counterparts were thus removed from the DH type-IV catalog.
\section{Results}
\label{sec:results}
{ The final DH type-IV sample contains 139 events, each of which is well associated with a white-light CME. This is in contrast to the case of metric type-IV bursts which can occur without a white-light CME associated~\citep{Anshu21_C24typIV_CME_corr}}.
The sample has more events than previously reported for the specific periods covered by previous works. For instance, in 1998 - 2012 period we report 91 events as opposed to 48 reported by \cite{Hillaris16_typeIV}. The rise in source count is partly because of the addition of data from STEREO spacecraft, unused in \cite{Hillaris16_typeIV}.
But, even if only the Wind/WAVES data within 1998 - 2012 are considered, our catalog has 71 events with a type-IV signature.
Similarly, the current list has 65 Wind/WAVES events in cycle 23, compared to the 42 events reported by \cite{Gopal11_PREconf}.
This rise in the source count can be attributed to the blind search strategy employed by combining date-time periods across metric type-IV list, DH type-II catalog and the LASCO CME catalog.
Also, the addition of STEREO data helped confirm some faint type-IVs in Wind/WAVES data which would have been omitted previously. Such faint events could then be confirmed based on the the detection of metric type-IV counterparts from ground based data. 

Now that the authenticity of events in our sample and completeness across earlier catalogs are verified, we put together a catalog of the characteristics of these bursts in the radio DS, properties of the associated CMEs and the flare location as recorded by multiple spacecraft. The latter gives a range of viewing angles for each event letting one explore the directivity of type-IV bursts.
The following sub-section describes the contents of the catalog.

\subsection{The DH type-IV catalog}
\label{sec:catalog}
\begin{figure}[ht!]
\epsscale{1.1}
\plotone{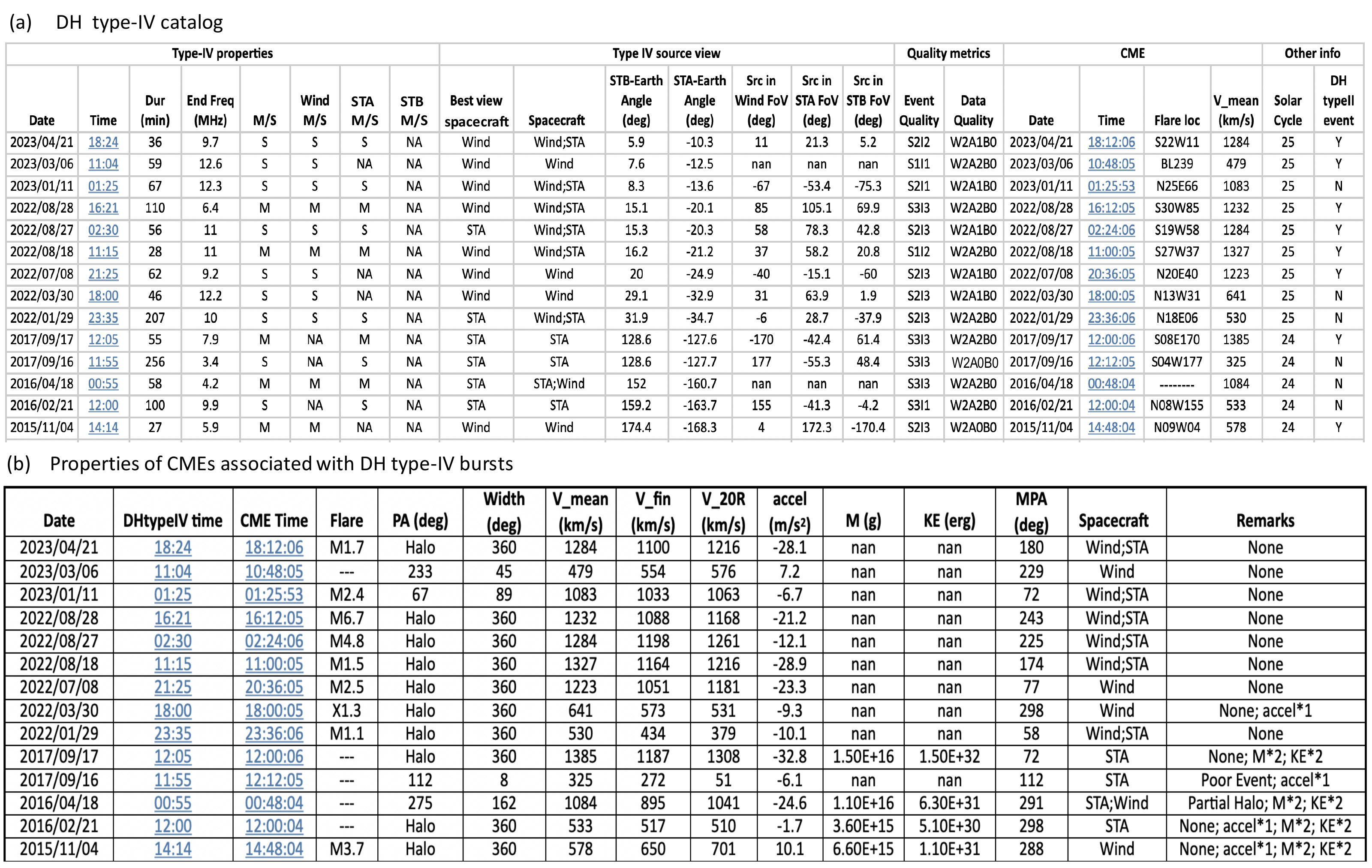}
\caption{(a): A portion of the DH type-IV catalog. (b): The corresponding table of properties of the associated CMEs.
\label{fig2:catalog}}
\end{figure}
\begin{figure}[ht!]
\plotone{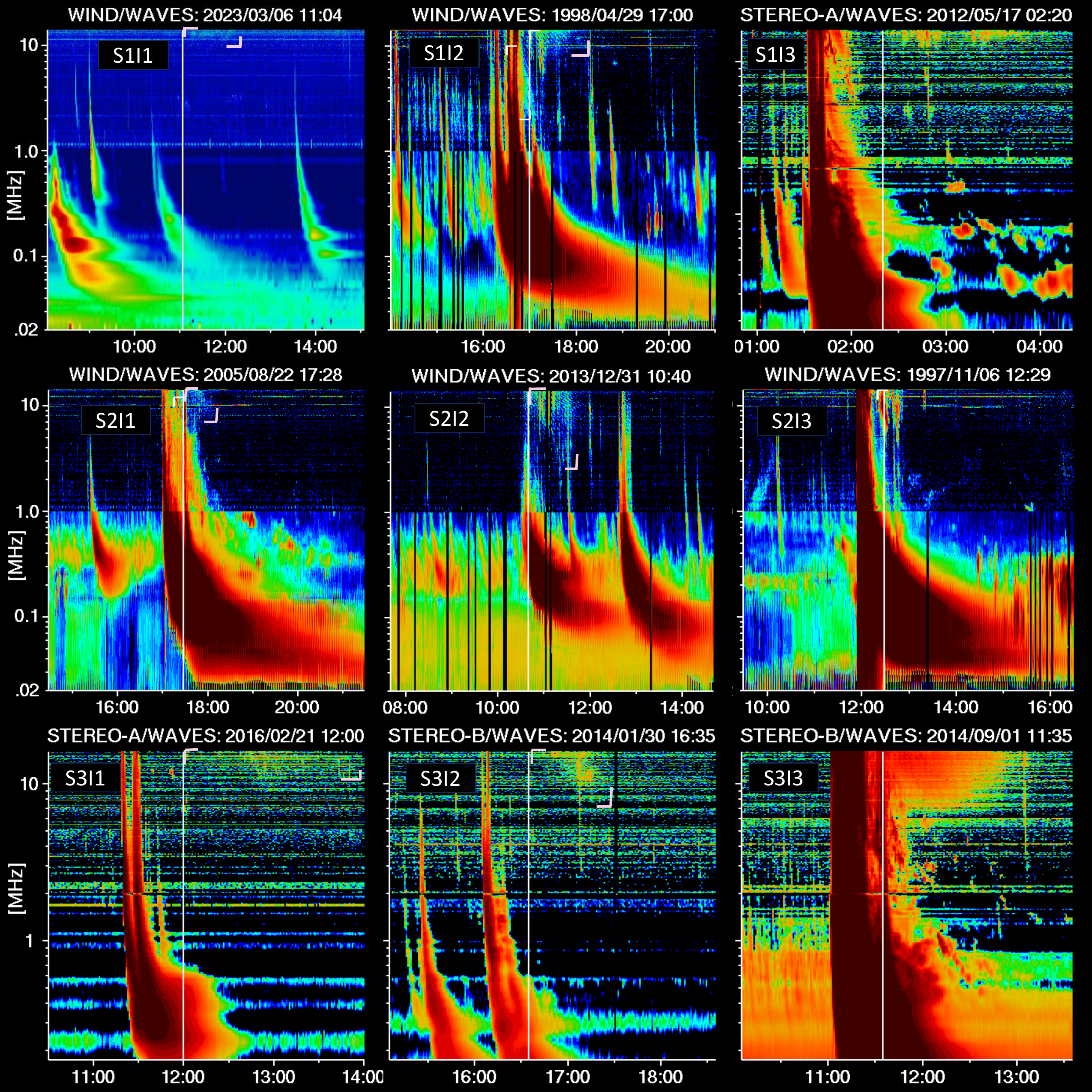}
\caption{Collage of DH type-IV bursts varying in event quality, SiIj where i,j$\in${1,2,3} with 1 being worst and 3 being best. S refers to the morphology and I refers to the intensity of the events.
\label{fig3:eventQlty}}
\end{figure}
\begin{figure}[ht!]
\plotone{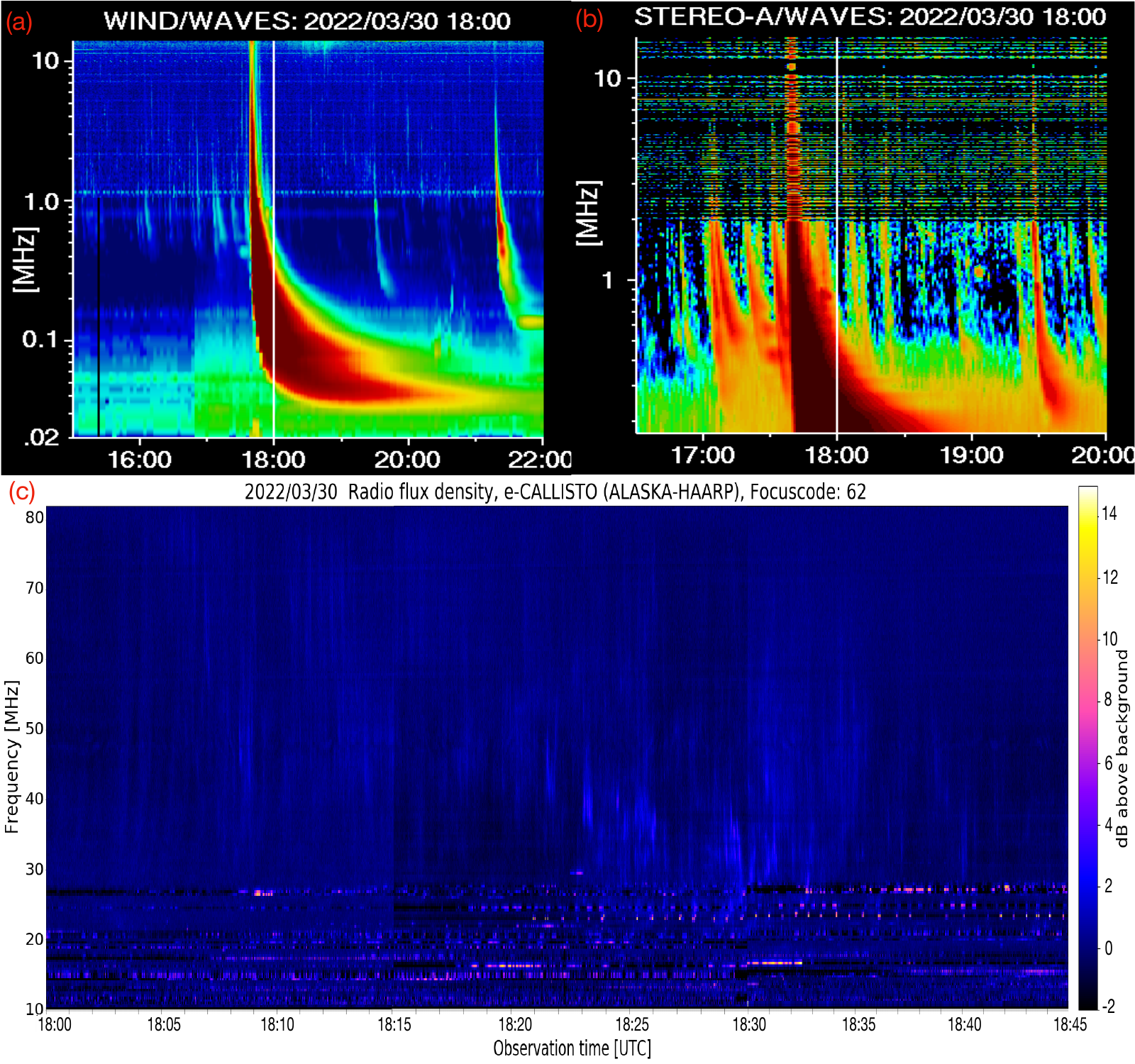}
\caption{An example event with a data quality metric of W2A1B0. (a-b): DH type-IV is clearly detected by Wind/WAVES in $>$10\,MHz between 18:00 - 19:00 UT on March 30, 2022 (W2); The quality of the STEREO-A data in 1 - 14\,MHz  is poor (A1), while STEREO-B was non functional during the period (B0). (c): A metric type-IV event recorded by e-CALLISTO station at Alaska. Data source: \href{http://soleil.i4ds.ch/solarradio/callistoQuicklooks/?date=20220330}{http://soleil.i4ds.ch/solarradio/callistoQuicklooks/?date=20220330} 
\label{fig3:dataQlty}}
\end{figure}
Figure~\ref{fig2:catalog}a shows a portion of the DH type-IV catalog, the full version of which is published online\footnote{\href{https://cdaw.gsfc.nasa.gov/CME\_list/radio/type4/DHtypeIV\_catalog.html}{https://cdaw.gsfc.nasa.gov/CME\_list/radio/type4/DHtypeIV\_catalog.html}}.
Cross matching with the LASCO-CME catalog provided the properties of the associated CMEs (\Vm, acceleration , kinetic energy, mass and width), forming an additional table shown in Fig.~\ref{fig2:catalog}b.
A description of the catalog, the CME characteristics table and the dynamic spectra are published online\footnote{\href{https://cdaw.gsfc.nasa.gov/CME\_list/radio/type4}{https://cdaw.gsfc.nasa.gov/CME\_list/radio/type4}}.
The DH type-IV catalog is broadly divided in to 5 sections, namely Type-IV properties, Type-IV source view, quality metrics, CME information and Other info.
The first section provides the minimum frequency of emission and the time duration of emission at 14\,MHz (referred to as duration hereafter).
In cases where multiple spacecraft report on an event, the lowest observed minimum frequency and the longest observed duration are reported in the catalog. This procedure also help us identify the spacecraft that recorded the `best view' of the type-IV event. For instance, consider the event in Fig.~\ref{fig1:example}a. The longest duration and the lowest minimum frequency is for the STEREO-A event which subsequently provided the best view of the burst compared to other spacecraft observations.
The first section also classifies events into either moving (M) or stationary (S) type-IVs across spacecraft, while also mentioning the M or S classification in the best view spacecraft (in `M/S' column).
The second section in the table provides details of the multi-vantage point observations. The spacecraft that recorded the `best view' is noted.
Also, all the instruments that detected each event is reported in the `Spacecraft' column. The relative positions of the spacecraft with respect to Earth is provided.
Using the CME flare location obtained from LASCO CME catalog, the relative source locations in the visible solar disk of STEREO-A, STEREO-B and Wind spacecraft are estimated for each event.
We use heliographic coordinates, so the source longitude ranges from -90$^\circ$ (east limb) to +90$^\circ$ (west limb) direction across the visible solar disk.
Thus the relative location of type-IV bursts observed by multiple spacecraft provide a database to systematically explore emission directivity over our larger sample.
The quality metrics will be introduced in detail in Sec.~\ref{sec:qlty}.
In the CME section of the table, the date, time, flare location and the mean CME speed estimated are provided based on LASCO CME catalog.
The `Other info' section gives information on the solar cycle and on the association of DH type-II bursts.
Combining all these information, we put together a comprehensive DH type-IV catalog for the period from 1996 to May, 2023, covering two full solar cycles and the rising phase of cycle 25. 
\subsubsection{Quality metrics}\label{sec:qlty}
The DH type-IV catalog reports two quality metrics for each event, namely `event quality' and `data quality'. 
`Event quality' defines the quality of the detected type-IV event along two axes namely, shape~(S) and intensity~(I). The events reported by the best view instrument are ranked based on the dynamic spectral shape and intensity along an integer scale from 1 to 3. Note that 1 is the worst and 3 is the best.
{ The ranking was done based on visual appearance. A type-IV burst is expected to appear as a broadband emission with a low-frequency cut-off. The cut-off can vary with time and the emission usually lasts for several minutes to hours. 
Based on how well the above features are traceable in the DS, the shape of the event is ranked. 
Since the burst morphology is often complicated in the frequency-time plane with multiple variability scales, defining a mathematical model is difficult. Also in many cases, the event is seen to be confined above 10\,MHz making the shape identification and ranking hard. This is why a manual analysis mode was opted for. Now that we have a manually characterized set, this could be used for building AI/ML-based burst identification models in the DH band.}
Figure~\ref{fig3:eventQlty} shows a collage of selected spectra with varying event quality metrics. The square brackets mark the type-IV burst emission. Consider the top row in the figure. Moving from I1 to I3 the intensity of the type-IV emission can be found to increase with respect to the background making it easily discernible.
However, note that the morphology of the emission is not well defined and these events are hence identified based on co-observations from other spacecraft and metric band DS from ground-based telescopes.
{ Direct observation of a post-flare loop that brightens up in co-temporal X-ray images was used as supporting evidence, increasing the likelihood of the emission being a type-IV, in cases of weak events.}
Now, consider the rightmost column in the collage. The relative intensity of the emission is very high (I3). As we move from top to bottom, the type-IV emission structure can be found to become clearer as the minimum frequency of the emission extends to lower frequencies in the observing window. This effect can be seen moving down the other two columns as well.
Note that the middle column shows different moving type-IV events.
In this scale, the event in Fig.~\ref{fig1:example}a middle panel would be ranked S3I3 due to the distinctive type-IV shape and intensity contrast, while the STEREO-B event would be S3I1.
However, the catalog would report event quality as S3I3 corresponding to the best detection. The spacecraft with the best quality detection is reported as the `Best view spacecraft' (see, Fig.~\ref{fig2:catalog}(a)).
{ We note that events with good `S' metric are usually the ones that extend below 10\,MHz in the dynamic spectrum and appear relatively isolated in the dynamic spectrum from other bursts, both aiding in detecting and characterizing them better. Often when a type-IV burst is not well isolated in the DS, despite its frequency extent or duration, it could get a `shape' metric of 2 rather than 3. A shape metric of 1 could suggest that the event was either in a very crowded part of the DS interlaced with several other bursts, or the event did not extend much in the frequency and time axes to discern the shape well.}

The `data quality' metric rates the quality of the DS data recorded by all spacecraft, during a burst event. The spacecraft are tagged by letters W, A and B for Wind, STEREO-A and STEREO-B, respectively. The metric is formed by combining the letter tag with an integer quality metric. { The integer metric is either 0, 1 or 2 denoting the absence of spacecraft data, the existence poor quality data or the availability of good quality data.  Fig.~\ref{fig3:dataQlty}a-b shows an event where a DH type-IV is detected by Wind in the high frequency channels. Meanwhile, the $\sim$ 2 - 14\,MHz data recorded by STEREO-A is of poor quality to make a robust conclusion. The source was at 63.9$^\circ$W in the STEREO-A field of view. STEREO-B was non-functional during the period. Fig.~\ref{fig3:dataQlty}c shows a simultaneous metric dynamic spectrum showing the event during the same period, from an e-CALLISTO station at ALASKA.
Meanwhile, the two events shown in Fig.~\ref{fig1:example} are ranked W2A2B2 since the data quality is good in all spacecraft irrespective of the detection of type-IVs. 
}
The two metrics are crucial for identifying reliable, good quality observations by multiple spacecraft, for a statistical study of the effect of line of sight in the observed radio burst characteristics.
However, to explore the mean statistical properties of DH type-IV bursts one can use the values in the Type-IV properties section which is by definition is based on the `best view' instrument data. 
\subsection{Characteristics of DH type-IV bursts and associated CMEs}\label{sec:sample_stats}
\begin{figure}[]
\epsscale{0.98}
\plotone{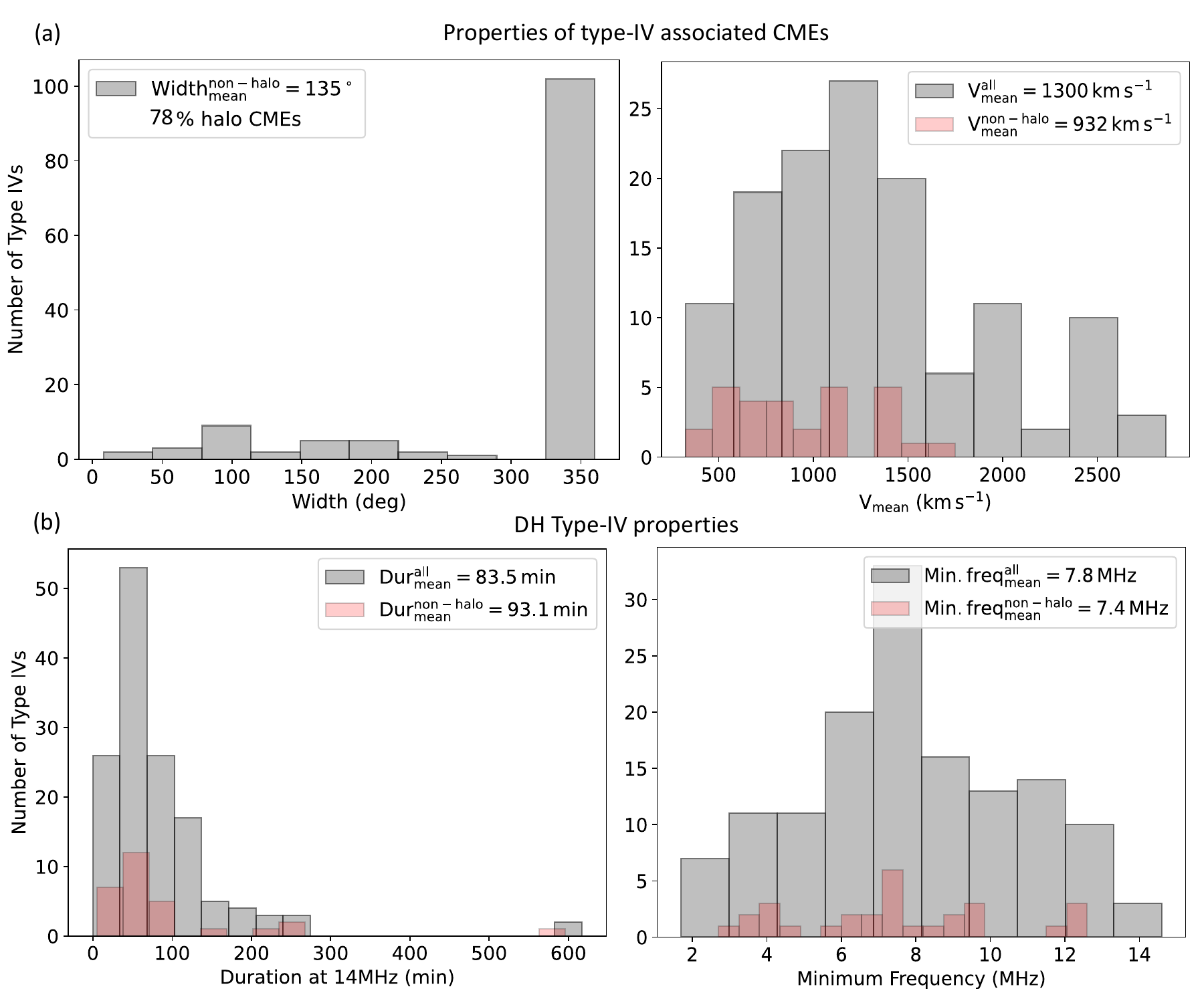}
\caption{{ (a): Properties of the CMEs associated with the DH type-IV bursts. 78\% of the events are halo-CME associated. (b): Properties of the DH type-IV bursts. Legends show the mean values of each property for all bursts and for those associated with non-halo CMEs. The mean properties of DH type-IV bursts caused by non-halo CMEs are similar to the full sample. But, the $\mathrm{V_{mean}}$ is slightly lower for the non-halo events.} 
\label{fig4:prop_hists}}
\end{figure}

Figure~\ref{fig4:prop_hists}a shows the histograms of minimum frequency and duration at 14\,MHz for all 139 events, while the bottom panel of the figure shows the mean velocity and angular width of the associated CMEs. Note that 8 events did not have good LASCO data, either due to data gap or saturation effects from the bright event, which reduced the number of events to 131 in our CME statistics. However, a CME is detected in all of these 8 events in the extreme Ultraviolet and X-ray images.

\begin{table}[]
\centering
\begin{tabular}{|c|l|ll|ll|}
\hline
\multirow{2}{*}{}                    & \multicolumn{1}{c|}{\multirow{2}{*}{\textbf{Property}}} & \multicolumn{2}{c|}{\textbf{All events}}             & \multicolumn{2}{c|}{\textbf{Non -halo CMEs}}         \\ \cline{3-6} 
                                     & \multicolumn{1}{c|}{}                                   & \multicolumn{1}{l|}{\textbf{Mean}} & \textbf{Median} & \multicolumn{1}{l|}{\textbf{Mean}} & \textbf{Median} \\ \hline
\multirow{2}{*}{\textbf{DH type-IV}} & Duration (min)                                          & \multicolumn{1}{l|}{83.8}          & 62              & \multicolumn{1}{l|}{93.1}          & 59              \\ \cline{2-6} 
                                     & Min.   frequency (MHz)                                  & \multicolumn{1}{l|}{7.8}           & 7.5             & \multicolumn{1}{l|}{7.4}           & 7.2             \\ \hline
\multirow{2}{*}{\textbf{CME}}        & Vmean ($\mathrm{km\,s^{-1}}$)                           & \multicolumn{1}{l|}{1301}          & 1223            & \multicolumn{1}{l|}{932}           & 874             \\ \cline{2-6} 
                                     & Width   (deg)                                           & \multicolumn{1}{l|}{310}           & 360             & \multicolumn{1}{l|}{135}           & 120             \\ \hline
\end{tabular}
\caption{{ Statistics of the various properties of DH type-IVs and associated CMEs. Characteristics of events associated with non-halo CMEs (width\textless{}300$^\circ$) are separately provided. 102 out of 139 events were non-halo CMEs.}}
\label{tab1:prop_stats}
\end{table}

\begin{figure}[ht!]
\epsscale{1.2}
\plotone{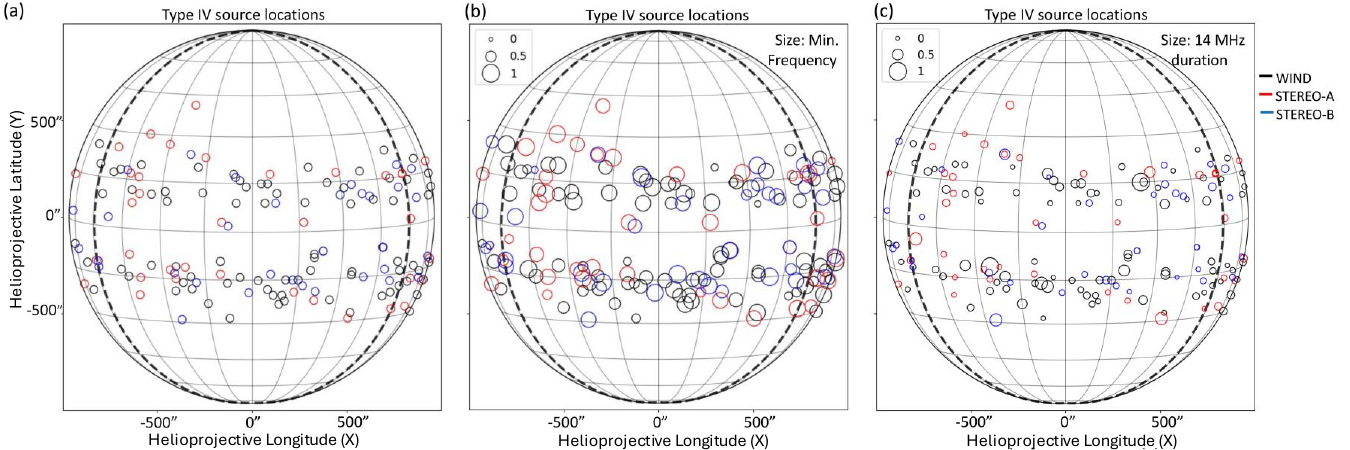}
\caption{{Stonyhurst maps showing the location of the active regions associated with DH type-IV bursts, in the field of view of each spacecraft. (a): Colors denote different spacecraft. 82\% of the type-IV sources lie within $\pm 60^\circ$ longitude, demarcated by the thick dotted great circles. The source latitudes, except in one case, are confined within $\pm 30^\circ$. (b-c): Same information as (a) with the marker size denoting the minimum frequency and the duration of the burst on a scale relative to the respective maximum value. Neither minimum frequency nor duration show a clear preference to source location.}
\label{fig5:srcloc}}
\end{figure}
Table~\ref{tab1:prop_stats} shows the statistics derived from the property distributions in Fig.~\ref{fig4:prop_hists}. The median values of duration and minimum frequency are 62\,min and 7.5\,MHz respectively, while the associated CMEs showed a median V$_{mean}$ of 1223\,km\,s$^{-1}$ with $\sim$ 78\% [102 out of 131] of them being halo CMEs~\citep[e.g.,][]{Howard82_HaloCMEDiscovery,Zhou03_HaloCMEs_surfaceactivityCorr, gopal07_geoeffectiveness_HaloCMEs}.
Halo CMEs are wide events seen extending around the entire occulting disk of the coronograph. { There is no significant difference in the mean properties of the DH type-IV bursts associated with non-halo CMEs and the larger sample. However, the mean CME speed of non-halo CMEs causing the DH type-IV bursts is around 932\,km\,s$^{-1}$ compared to the full sample mean of 1301\,km\,s$^{-1}$.} 
Meanwhile, halo CMEs can be a disk centre event, a limb event or a backside event~\citep{Gopal04_CMEbookChapter}. 
However, since the post-flare loops and moving plasmoids in backside events will be mostly occulted by the disk, only the DH type-IV bursts associated with front-side or limb events are observable.
Figure~\ref{fig5:srcloc}(a) shows the distribution of DH type-IV sources on the solar disk as seen by multiple instruments. The dashed bold longitudes mark the $\pm$60$^\circ$ region. About 82\% of the sources are concentrated within this region, demonstrating a strong directivity in the longitude distribution of these events. 
{ The source latitudes are clearly within $\pm 30^\circ$, indicating that the underlying CMEs originate from the active region belt. Only active regions have the ability to produce powerful CMEs that produce DH type-IV bursts.
The other panels in Fig.~\ref{fig5:srcloc} show the same information on the source distribution as, but the marker sizes represent minimum frequency (b) and the duration of the burst. The sizes are denoted on a scale relative to their respective maximum property value. The size distribution has no clear dependency on the latitude or longitude of the burst sources.}
%However, no clear pattern is seen in the latitudinal distribution of the minimum frequency and burst duration.
\begin{figure}[ht!]
\epsscale{1}
\plotone{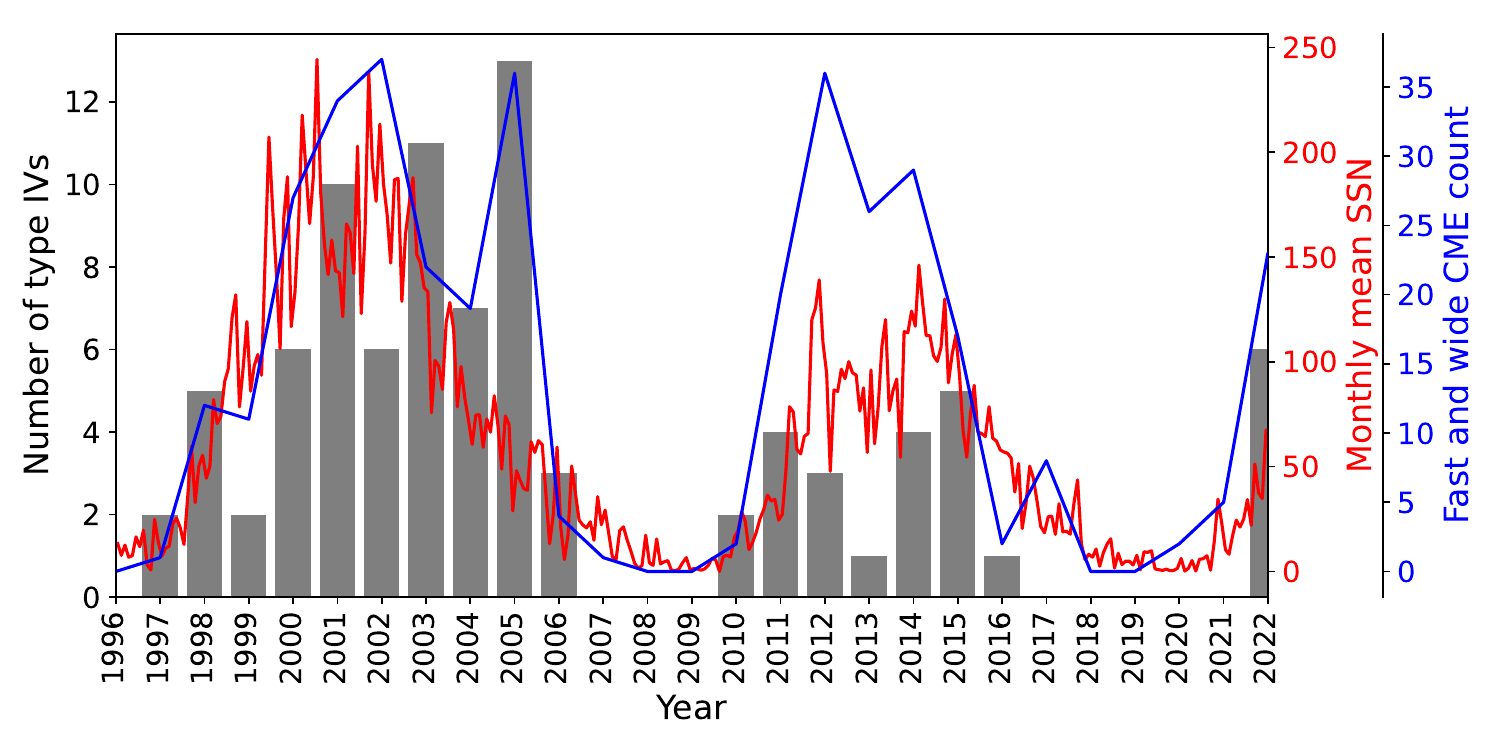}
\caption{Solar cycle variation in the number of DH type-IV bursts observed by Wind/WAVES. Monthly mean sunspot number (red) and the number of fast and wide CMEs (blue) are shown. 73\% of the DH type-IV bursts in the catalog are associated with a fast and wide CME.
\label{fig7:srccount_cycle}}
\end{figure}
\section{Discussion}\label{sec:discussion}
%\cite{gopal07_geoeffectiveness_HaloCMEs} did a statistical study of such events for cycle 23 and showed that 71\% of the frontside halo events (limb and disk center) are geoeffective, with the fraction close to 75\% for disk halos.
%The authors showed that disk halos generally cause intense geo-magnetic storms with Dst indices dropping below -100\,nT.
%This makes DH type-IV events well connected to strong geo-effective events.
The mean CME and radio burst properties listed in Table.~\ref{tab1:prop_stats} agrees with the previous results by other authors within a few \%~\citep{Gopal11_PREconf, Hillaris16_typeIV}, except for \Vm.
\cite{Gopal11_PREconf} reported a mean \Vm\ of 1526\,km s$^{-1}$, compared to our value of 1301\,km s$^{-1}$. 
Also, we find a 32\% association of DH type-IVs with CMEs having \Vm$<$1000\,km s$^{-1}$, as opposed to 20\%~\citep{Gopal11_PREconf} and 26\%~\citep{Hillaris16_typeIV} association reported earlier.
It is possible that our blind search criteria combining multi-spacecraft data helped nearly triple the number DH type-IV bursts, with more events associated with weaker CMEs providing a more statistically complete dataset.
However, we broadly agree with the inferences of the earlier publications that DH type-IV bursts typically last for over an hour, extending on average to frequencies $\approx$7.5\,MHz and are commonly associated with fast ($>$900\,kms$^{-1}$) and wide ($>$60$^\circ$) CMEs. 96 out of 131 bursts (73\%) with a well identified source location, have a fast and wide CME association. Also, 78\% (102 out of 131) of the bursts are associated with a halo CME.
{ We explored the non-halo CMEs associated with DH type-IV bursts separately. The mean speed and the width of these CMEs also fall well within the definition of a fast and wide CME.}
%\citep{Gopal11_PREconf,Hillaris16_typeIV}
Figure~\ref{fig7:srccount_cycle} shows the solar cycle variation in the DH type-IV detection count from the vantage point of the Earth, as seen by Wind/WAVES. The number of type-IVs clearly follow the variation in the number of fast and wide CMEs. Though it follows the monthly mean sunspot number, the correlation is stronger with the number of fast and wide CMEs. This is evident from the fact that the type-IV counts even follow the bump around 2005 - 2006 period and the dip around 2013 seen in the fast wide CME counts.
However, despite this strong generic association between two, the number of fast and wide CMEs is much higher than that of the DH type-IV bursts.
In fact, within Nov 1996 to May 2023, only 18\% of the fast halo CMEs produced a DH type-IV burst. 
This means that even if a CME with the most favourable physical attributes occur in terms of speed and angular extent, the odds for an associated DH type-IV burst is fairly low.
{ Besides, DH type-IV bursts show a 67\% (93 out of 139 events) association with DH type-IIs.}

\begin{figure*}[ht!]
\epsscale{0.99}
\plotone{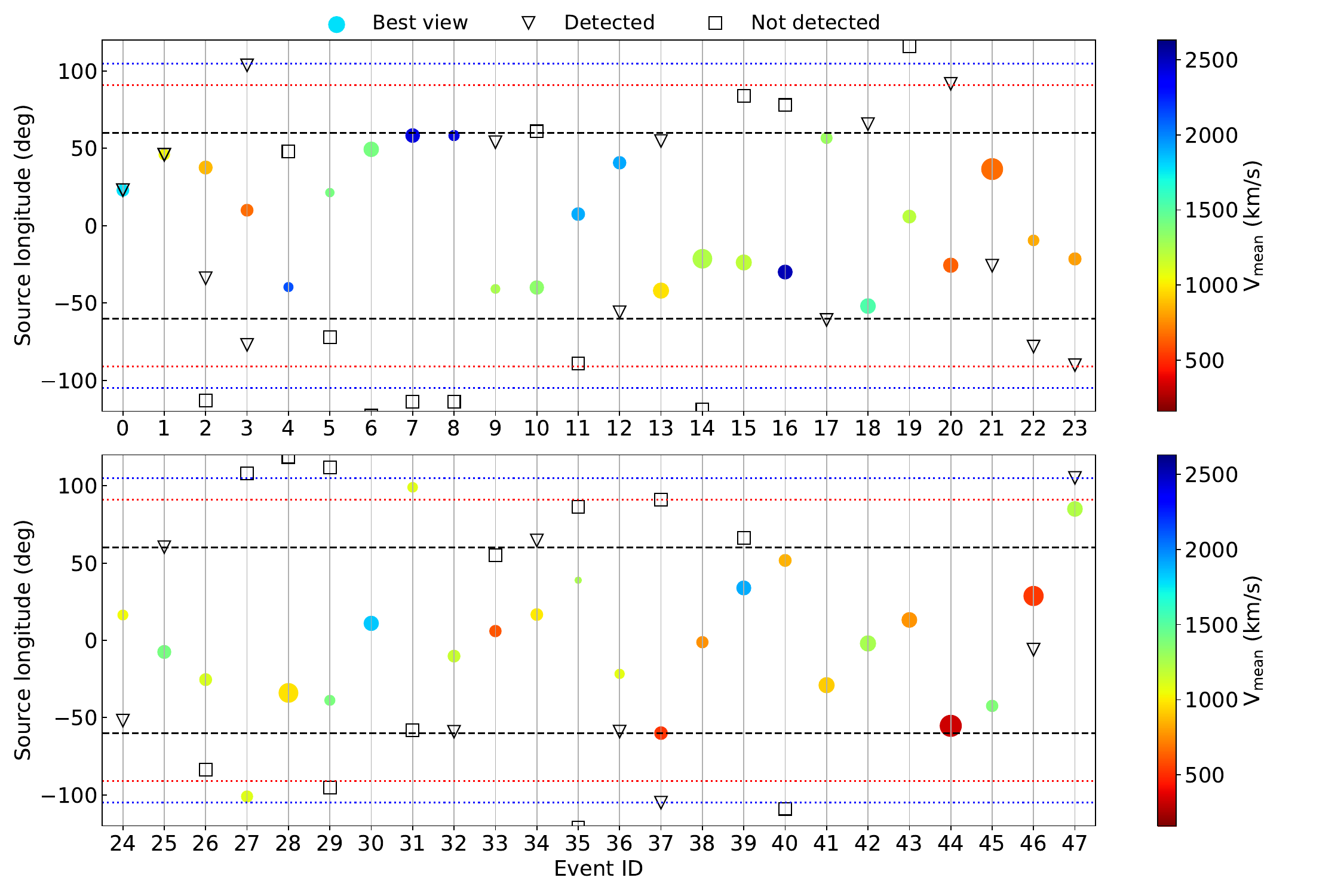}
\caption{{ All 48 events that had good observations from at least 2 spacecraft. X-axis shows a representative event ID for each event.} Y-axis shows the burst source longitude relative to the spacecraft vantage point. The relative source longitudes that gave a detection are marked by triangles and circles, with circles marking the best detection. The size of the circle represent the burst duration while the color denote \Vm. Dashed horizontal lines mark $\pm$60$^{\circ}$ $\pm$90$^{\circ}$ and $\pm$105$^{\circ}$ longitudes. { Longitude $>|90|$ represent flare sources behind the limb. In some cases where the post-flare loop rose much higher, a type-IV emission could be detected from flare sources slightly behind the limb.}
\label{fig6:srcloc_multisat}}
\end{figure*}
\begin{figure*}[]
\epsscale{0.99}
\vspace{-0.5cm}
\plotone{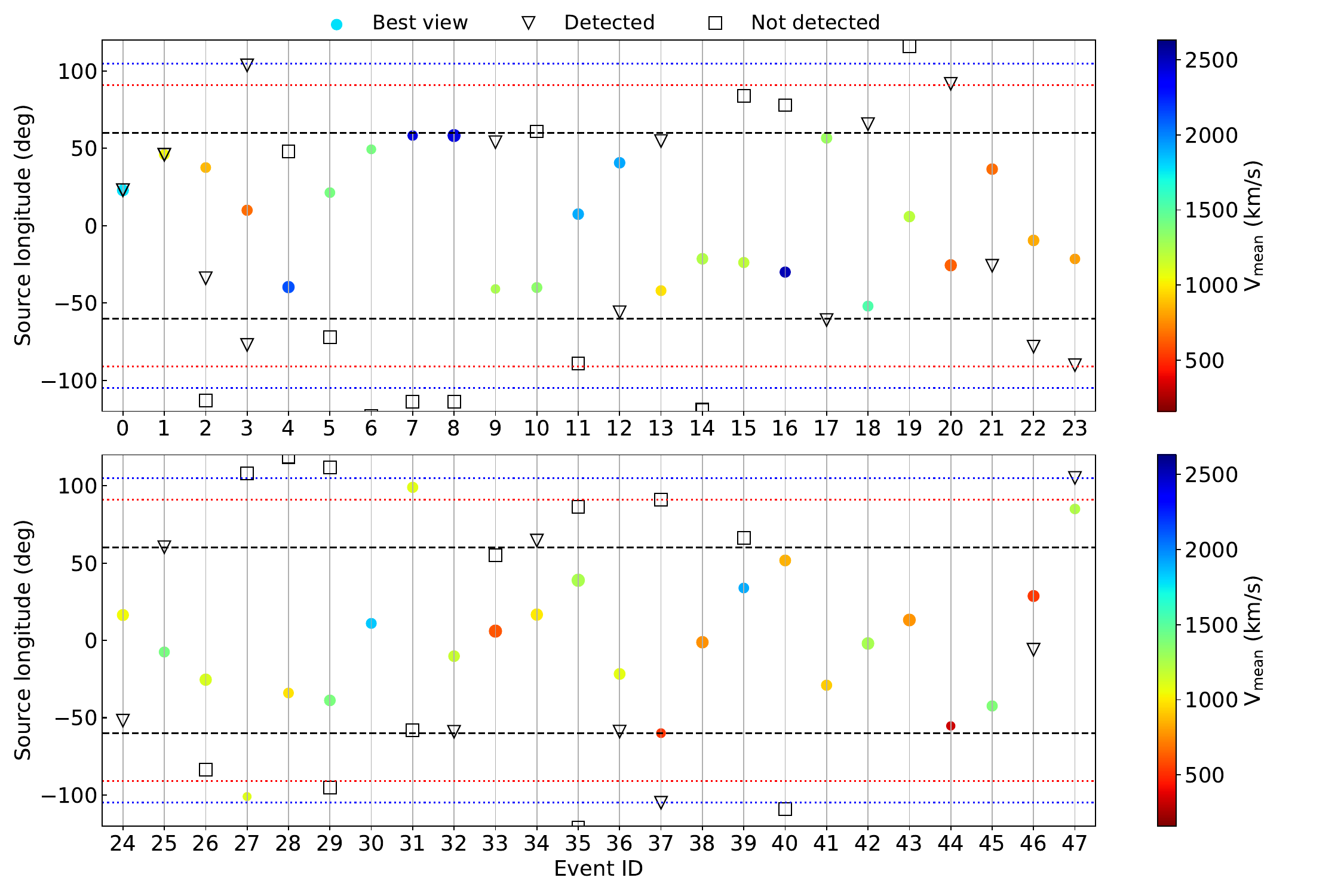}
\caption{Same as Fig.~\ref{fig6:srcloc_multisat}. Sizes of the circles represent minimum frequency.
\label{fig6:srcloc_multisat_1}}
\end{figure*}
\begin{figure*}[]
\epsscale{1}
\plotone{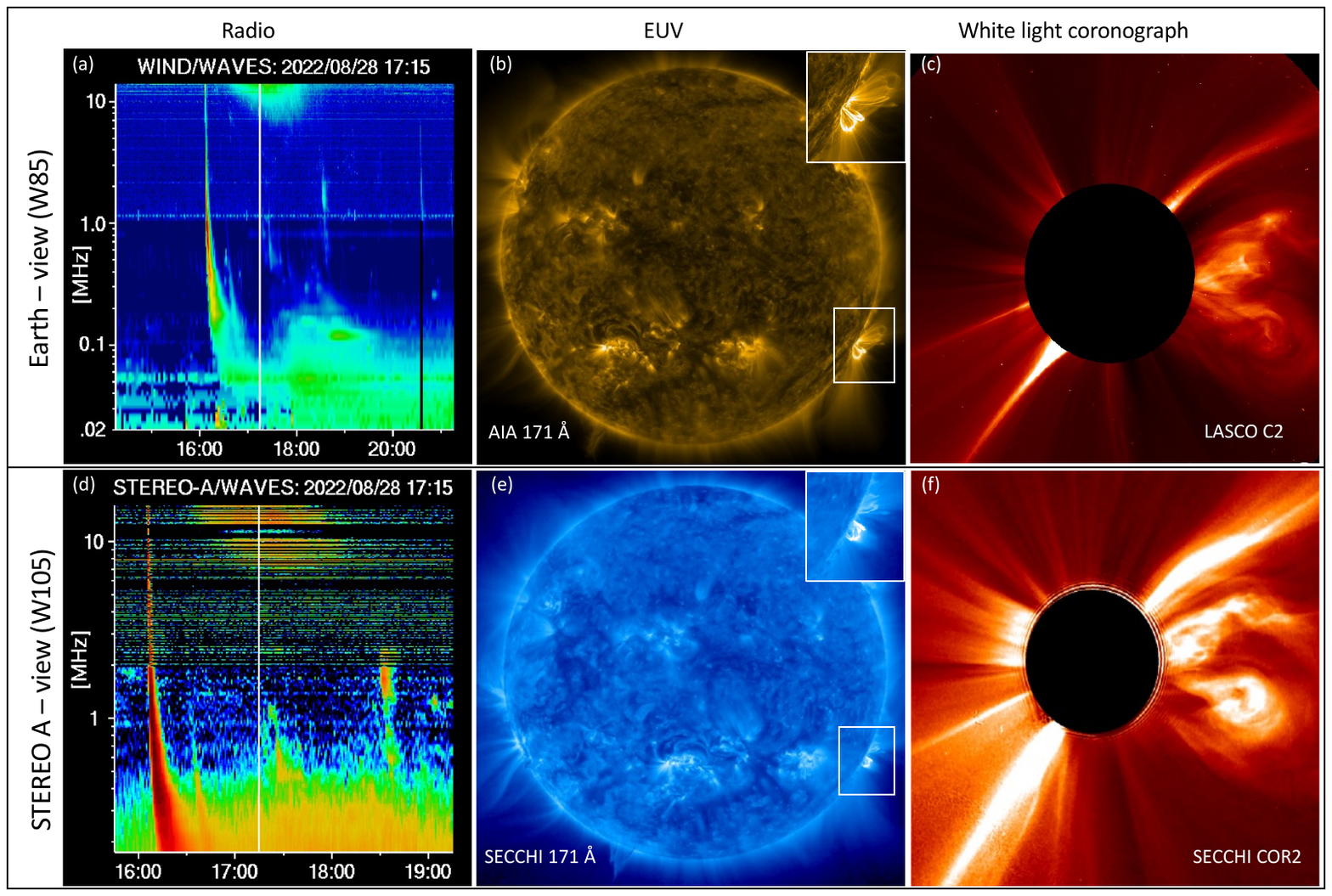}
\caption{{{ Case of a type-IV from a flare source behind the limb (Event 47):} Radio, EUV and white-light coronograph observations of a limb CME event from the view points of the Earth (top row) and STEREO-A (bottom row). Source locations are specified in the left of each row (Wind: 85$^\circ$W; STEREO-A: 105$^\circ$W). STEREO-A was at 20$^\circ$E with respect to the Earth. DH type-IV burst was observed by both spacecraft. Associated post-flare loop is visible over the limb in the EUV maps from both vantage points. The insets in the middle panel zooms in to the loop region. The rightmost panels show the CME in the coronagraphs from the Earth and STEREO-A vantage points}.
\label{fig8:ex_limbevent}}
\end{figure*}
\begin{figure*}[]
\epsscale{0.98}
\plotone{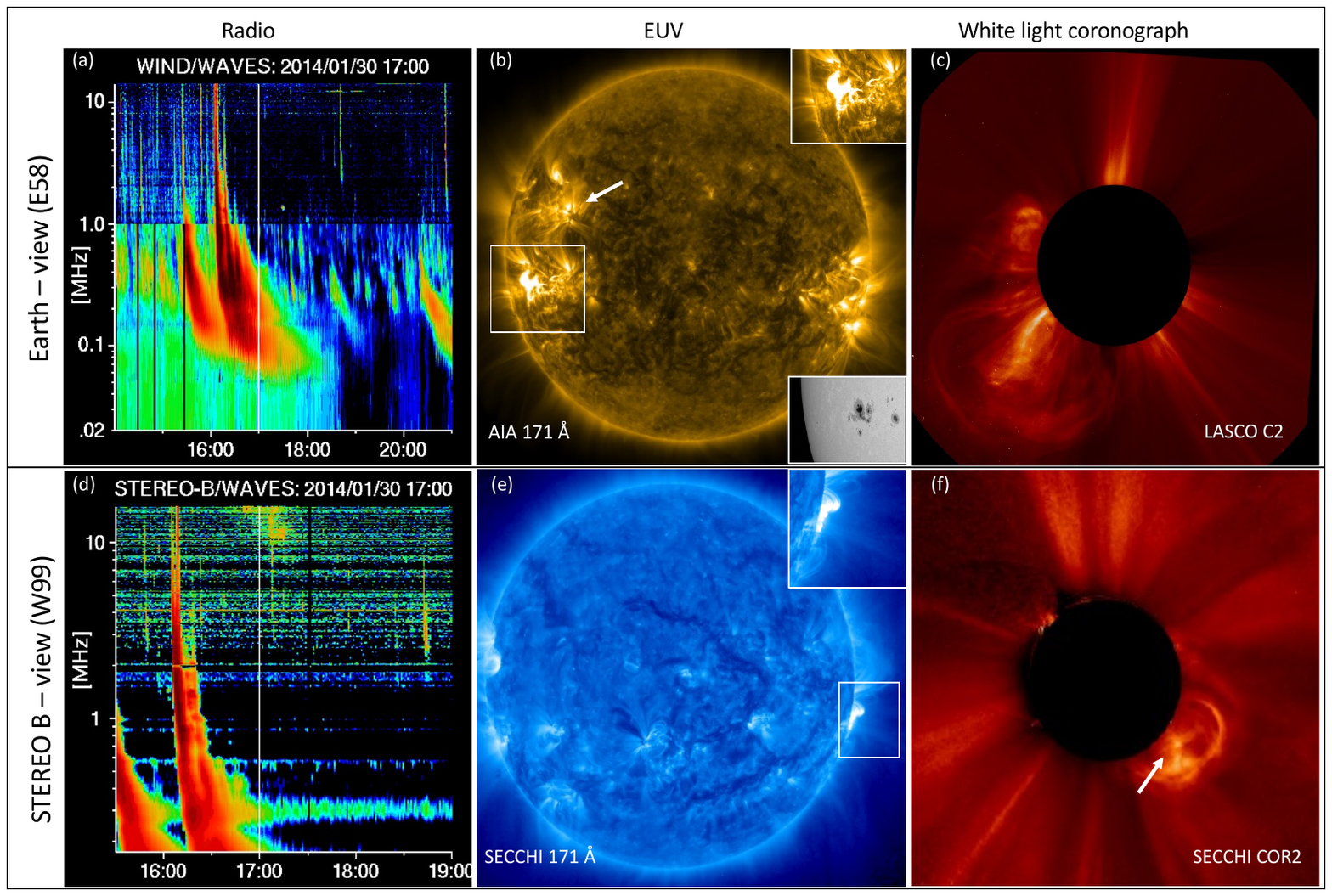}
\caption{
Case of { Event 31}: Radio, EUV and white-light coronograph observations of the CME from the view points of the Earth (top row; source longitude: 58$^\circ$E) and STEREO-B (bottom row; source longitude: 99$^\circ$W). STEREO-A was at 150$^\circ$W with respect to the CME source. DH type-IV burst is detected by STEREO-B but not Wind. Insets in the top right region of the middle panels (b and e) zooms in to the white boxes that mark the flaring region. In panel b, a HMI image of the flaring region is provided in bottom right inset. { The arrow in panel b marks the nearby active region while in panel f, the arrow points at the CME plasmoid temporally correlated with the observed DH type-IV. The active region marked in panel b is the likely source of the noise storm in Wind/WAVES in the pre-flare period (before $\sim$ 16:00 UT). STEREO-B does not detect this emission possibly because of the directivity weakening the flux density of the noise storm source which is only partly visible at the limb in STEREO-B field of view.}
\label{fig9:event29}}
\end{figure*}

\subsection{Longitude of type-IV burst sources - emission directivity}
As mentioned earlier, \cite{gopal16_typeIVdirectivity} showed that the type-IV burst sources appear to be concentrated within $\pm$60$^{\circ}$ longitude. However, this was based on type-IV burst location distribution derived from cycle 23, a period before the STEREO era.
So also they did not have a sample of multi-vantage point observations of type-IV bursts to verify the directivity in a systematic manner.
However, the authors presented a single case study of an event from 2013 Nov 7 that was observed with the Wind and STEREO spacecraft. 
The DH type-IV event was found to be best observed by the spacecraft which was viewing the event within $\pm$60$^\circ$.
Based on this the authors argued that there could be a directivity in the inherent emissivity, besides any effect due to background plasma opacity.
A few case studies that followed this paper, argued that occultation by dense overlying material, often a CME-streamer shock formed towards the CME flank region, could be prime reason for the blocking the emission along certain viewing angles and not necessarily inherent emissivity~\citep{Melnik18_typeIVBehindLimb, Nasrin18_DHtypeIV_occult_streamerCMEshock,Pohjo20_typeIVdirective}.
Thus the earlier studies that explored directivity in the emission have either been on specific cases or on the distribution of observed source locations from one spacecraft without factoring in the variation in the dynamic spectral morphology detected from multiple vantage points.
The DH type-IV catalog presented here, has event and data quality metrics which grades the characteristic detail in the observed DS of each DH type-IV event for each spacecraft located at different vantage points. 
Based on these metrics, the catalog also mentions the best-view spacecraft which detected the type-IV burst with the best extent in frequency and time of all recorded data from multiple view points. This let us study the direction dependency in not only the detection of the DH type-IV burst, but also its morphological details in the DS.

To systematically explore the effects of viewing angle, the bursts detected with a good event quality of SiIj, where i,j$\in$\{2,3\} were first chosen. From these events, only those that occurred when at least two spacecraft were producing good dynamic spectra with data quality of 2 were chosen. This produced a subset of { 48 events with good multi-vantage point observations of the Sun from at least two viewpoints. The list of events with burst and CME characteristics and assigned event IDs are provided online\footnote{ \href{https://cdaw.gsfc.nasa.gov/CME\_list/radio/type4/All_good_multiview_events.html}{https://cdaw.gsfc.nasa.gov/CME\_list/radio/type4/All\_good\_multiview\_events.html}}.}
The event ID simply tags each event.
Figure~\ref{fig6:srcloc_multisat} shows the expected source longitudes of the selected DH type-IV burst sources (i.e, flare loops) in the frame of each spacecraft. Locations of the best-observed burst events are marked by colored circles, other detected source longitudes by triangles, and non-detections by squares. The non-detection longitudes provide the source location of the associated flare in the reference frame of the spacecraft which failed to detect it.
Marker colors of the best-viewed detections denote V$_\mathrm{mean}$ and sizes denote the type-IV duration. Figure~\ref{fig6:srcloc_multisat_1} is similar to Fig.~\ref{fig6:srcloc_multisat} with the exception that the sizes denote minimum frequency.
First, the role of viewing angle (or apparent source longitudes) in the detection of DH type-IV bursts is explored.
We find that { 58 out of 62 observations (94\%)} made within $\pm$60$^\circ$ source longitude by various spacecraft detected a burst (see, circles and triangles).
Meanwhile, only { 9 out of 20 observations (45\%)} made between $\pm$60$^\circ$ and $\pm$105$^\circ$ were detections. { Of the total type-IV detections reported when the spacecraft are viewing the source at an angle within $\pm 105^\circ$, only 13\% (9 out of 67) is contributed by spacecraft view points $>|60^\circ|$.}
An upper limit of 105$^\circ$ is set on the source longitude because of a few type-IV detections associated with flares that occurred slightly behind the limb.
The post-flare loops or plasmoids that caused the radio burst in these events extended sufficiently beyond the limb making them visible, as evidenced by the associated extreme Ultraviolet (EUV) movies.
For instance, Event { 47} was detected by two spacecraft, both viewing it close to the limb with flare source longitudes of 85$^\circ$ and 105$^\circ$. 
Figure~\ref{fig8:ex_limbevent} shows the DS for the event along with EUV and white-light coronograph images from the two spacecraft that detected it.
The EUV and coronograph images correspond to the time instants marked by the lines in the DS.
A post-flare loop can be found rising beyond the limb in the EUV images.
This implies that the DH type-IV bursts are detectable even beyond the the limb ($>90^\circ$ source longitude), provided the post-flare loop is visible over the limb and there are no dense occulting structures along the line of sight.
{ From Fig.~\ref{fig6:srcloc_multisat}, only 3 detections are reported out of the 14 such observations from $\gtrsim 90^\circ$ source line of sight.
%Even for lines of sight within $60 - 90^\circ$, only 3 out of 8 were detections.
} 
%However, the detection rate is significantly low for events beyond $\pm$60$^\circ$ compared to that in the complementary range.

Now we explore the effect of viewing angle in the observed morphology in the radio DS. 
From Fig.~\ref{fig6:srcloc_multisat}, it can be seen that for every event the spacecraft which recorded the best view (marked by circle) was always within $\pm$60$^{\circ}$ longitude, provided there was a spacecraft observing the source within that range.
The only exception for this rule is Event { 31}, which was hence studied in detail.
Figure~\ref{fig9:event29} shows the observations of the event in radio, EUV and white-light coronograph as seen from the viewpoints of the Earth (Top row) and STEREO-B (Bottom row). A DH type-IV burst was detected by STEREO-B but not by Wind, despite the latter viewing the source at $\sim$60$^\circ$. Meanwhile, for STEREO-B this was a limb event.
The post flare loop is clearly visible at the limb from the STEREO-B viewpoint as seen in the 171\AA\ image from SECCHI instrument on board STEREO-B. The inset of Fig.~\ref{fig9:event29}(e) shows a zoomed in image of the region.
However, the 171\AA\ image of the flaring region as seen from Earth, taken by AIA onboard SDO, reveal a complex active region configuration (Fig.~\ref{fig9:event29}(b)). The insets in the panel zooms into the flaring active region in 171\AA\ (top right) and HMI continuum (bottom right) maps. The HMI map reveals a complex $\beta\gamma$ sunspot group near the flare site. Also, another active region can be spotted (marked by arrow) within 10$^\circ$ separation of the one in question.  
Hence, given the complex magnetic field configuration in the neighborhood of the flare site, the radio bursts are likely occulted by the dense plasma in the complex post eruption field structures. 
The effect of source occultation on type-IV burst detection has been discussed in earlier works~\citep[e.g.][]{Nasrin18_DHtypeIV_occult_streamerCMEshock,Pohjo20_typeIVdirective}. 
This occultation effect could also be the cause of the disruption of the type-I noise storm in the Wind DS after $\approx$ 16:20 UT, when the CME associated type-III burst occurred.
Meanwhile, the type-IV burst recorded by STEREO-B is a moving type-IV and starting around 16:50 UT when the plasmoid structure marked by white arrow in Fig.~\ref{fig9:event29}(f) first appears in the SECCHI COR2 field of view.
%Probably, the moving plasmoid became visible in STEREO-B line of sight once it crossed the dense lower lying plasma. 

{ The inferences from the overall statistics and the special case of Event 31 suggest that in the absence of dense structures along the line of sight that can occult the view, a DH type-IV burst is best viewed when the spacecraft is within $\pm$60$^\circ$ LOS. However, DH type-IV bursts can be detected even from limb sources when there is no significant LOS occultation (e.g., Event 47), though this need not be the line of sight that provide the best view of the event.}
This is as expected from the hypothesis of \citep{gopal16_typeIVdirectivity} that the type-IV emission could have inherent directivity in the emission making it best viewed within $\pm$60$^\circ$ line of sight out of all possible detections in different lines of sight.

\subsection{Relations between DH type-IV properties, associated CME and viewing angle}
\begin{figure*}[]
\epsscale{1.12}
\vspace{-0.5cm}
\plotone{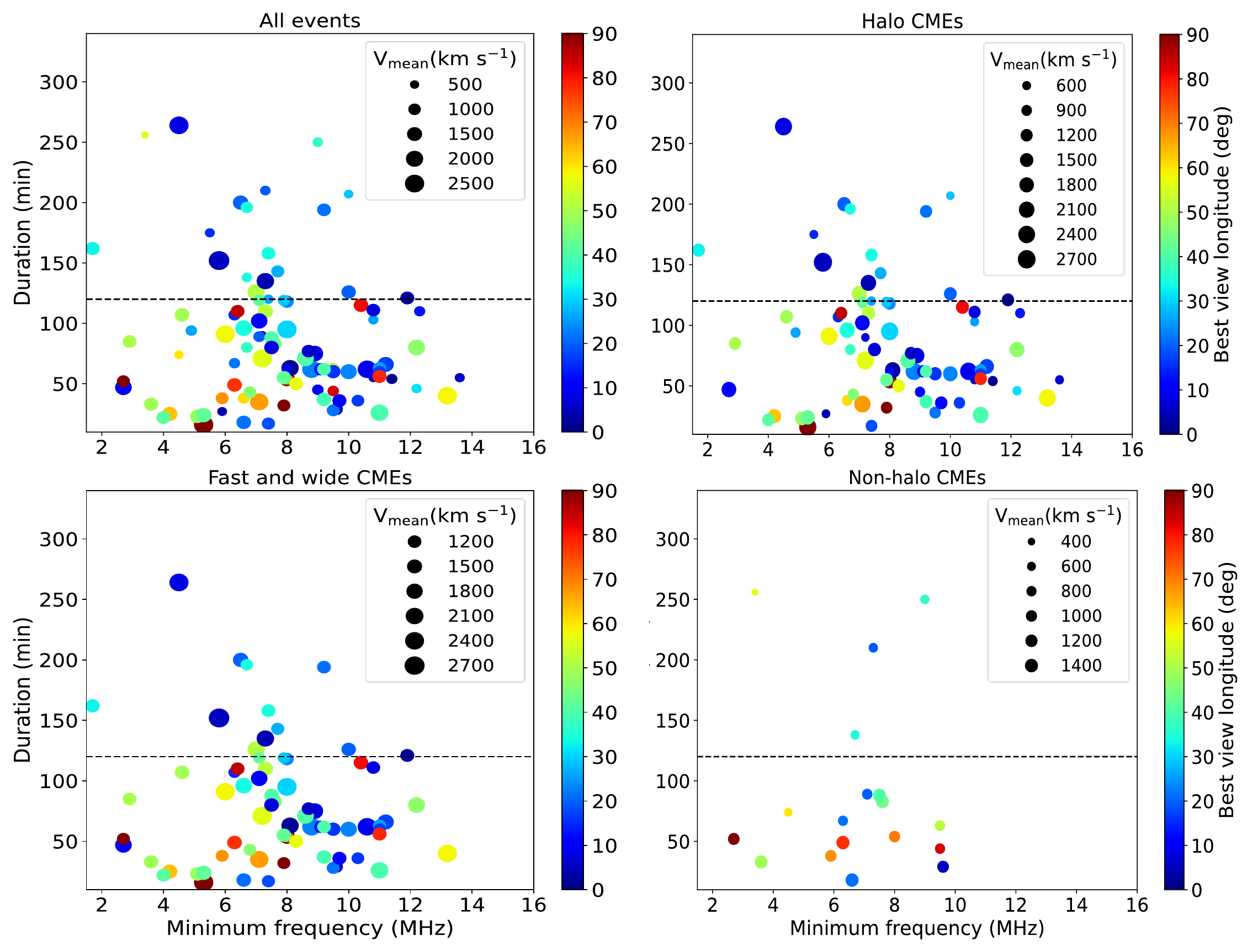}
\caption{Properties of DH type-IV events with quality of at least SiIj, where i,j$\in${2,3}. Modulus of source longitude is shown in color scale, while sizes denote \Vm. Horizontal line marks 120 min.\label{fig10:cluster}}
\end{figure*}
Since the viewing angle or equivalently the observed source longitude is a crucial factor in obtaining the best view of the burst, it is possible that the properties of the radio burst in the DS might depend on source longitude.
But, the colors and sizes of the circles in Fig.~\ref{fig6:srcloc_multisat} and Fig.~\ref{fig6:srcloc_multisat_1} do not show any particular pattern in their distribution across the longitude axis.
Since the colors denote \Vm, this means that the CME speed is not a crucial determinant in deciding the detectibility of a type-IV burst even close to the limb.
Similarly, the duration (in Fig.~\ref{fig6:srcloc_multisat}) and minimum frequency (in Fig.~\ref{fig6:srcloc_multisat}) of the radio burst denoted by the sizes, also do not show any definitive trend across longitude axis. 
This means that the emission directivity may not have a major impact on the observed DS characteristics.
We further explored the patterns and inter-relations, if any, in the multi-parameter space of the properties of the observed type-IV burst, associated CME and viewing angle. Only events with event quality, SiIj such that i,j$\in${2,3} were chosen for this study.
Figure~\ref{fig10:cluster} shows the variation in \Vm, type-IV duration, minimum frequency and best-view source longitude for all selected events. { We use the absolute value of the source longitudes since the effect of viewing angle should be symmetric about the central meridian.} Different panels explore the event characteristics for halo, non-halo and fast-wide CME associated events. Halo and non-halo CMEs were separately studied because the CME width histogram in Fig.~\ref{fig4:prop_hists}(b) shows a bimodal distribution. Also, the fast-wide CMEs were explored separately since their occurrence rates are seen to correlated well with that of DH type-IV counts (see, Fig.~\ref{fig7:srccount_cycle}).
The detection of DH type-IV bursts with duration longer than 120\,min (horizontal black line) seems to prefer a source longitude within $\pm$60$^\circ$.
A more detailed clustering analysis is currently underway by expanding the CME property table (Fig.~\ref{fig2:catalog}(b)) with more physical characteristics of the CME and associated flare. This is beyond the scope of this paper which is aimed at presenting the catalog, statistics, type-IV directivity and preliminary inferences on clustering of various properties of these events.

%No evidence for any sort of significant impact with viewing angle is found, though one could argue that long duration events generally prefer closer to disk center. 
% To do
% Plot vmean, Freq min, view ang color ; vmean , duration, view angle color

\section{Conclusion} %%%%%%%%%%%%%%
  \label{sec:conclusions}
We present a comprehensive catalog of Decametric Hectometric (DH) type-IV bursts during the period between Nov, 1996 and May, 2023, covering two complete solar cycles (23 and 24) and the rising phase of the current cycle. The catalog is complete to the entire LASCO CME event list and also covers the period of reported metric type-IV and DH type-II bursts. { All DH type-IV bursts are associated with white-light CMEs, in contrast to the case of metric type-IV bursts.}
Also, we include radio data gathered by three spacecraft namely Wind, STEREO-A and STEREO-B.
This not only help increase the source count, but also provides multi-vantage point observations for several DH type-IV bursts letting us explore the intriguing aspect of emission directivity.
The final catalog has 139 DH type-IV bursts, along with the properties of the associated CMEs. Since DH band dynamic spectrum is a disk integrated observable, the presented catalog form a sun-as-a-star database to explore the CME-related type-IV bursts, which are one of the much sought after radio signatures of stellar-CMEs. 

DH type-IV bursts are mostly associated with halo CMEs (78\%; 102 out of 131 events) with a mean speed of 1301\,km\,s$^{-1}$.  93 out of 139 events had a DH type-II association.
The solar source longitudes of the detected events lie within $\pm$60$^\circ$ for { 58 out of 62 cases} where multiple spacecraft observed the same source from different vantage points.
{ The source latitudes of all except one type-IV source is found to be within $\pm 30^\circ$ latitude pointing at the association of strong CME causing active regions with DH type-IV bursts. The sole outlier is however within $\pm 45^\circ$}
Our statistical results over a sample thrice larger than previous studies confirm the earlier results by different authors~\citep[e.g.][]{Gopal11_PREconf,gopal16_typeIVdirectivity} in a statistically robust manner.
We find a strong correlation between the occurrence of these bursts with that of the fast and wide CMEs during the catalog period.
However only 18\% of the fast halo CMEs that occured during the catalog period produced a DH type-IV burst. { So a DH type-IV detection indicates a high likelihood for the occurrence of a fast halo CME, but the converse is not true.}

Using the multi-vantage point observation data from Wind and STEREO spacecraft, the effect of directivity on the detection of these bursts and on their observed morphology in dynamic spectrum (DS) were explored.
When { 94\%} of cases where the spacecraft was observing a flare source within $\pm$60$^\circ$ line of sight, ended up in a detection, only { 45\%} of the observations made outside this field of view reported a burst.
In cases where multiple spacecraft observed an event, it is found that the spacecraft which observed the burst source within $\pm$60$^\circ$ longitude recorded the burst with the longest duration and extending down to the lowest frequency in the DS whereby providing the best or most detailed view of the event.
This rule had only one exception in the entire catalog, in which case we find evidence for significant occultation of radio emission by surrounding magnetic field structures. { However, in the absence of significant line of sight occultation, type-IV bursts from even limb events can be detected, though these detections may not provide the best possible view of the event.}
Additionally, the detection of a burst with duration $>$ 120\,min strongly prefer a viewing angle within $\pm$60$^\circ$.
Based on these results we infer that, though detecting a DH type-IV is possible at source longitudes extending up to the limb, a comparison of the dynamic spectra for the same burst as seen by multi-vantage point missions do suggest a inherent directivity in the radio emission that prefer $\pm$60$^\circ$ viewing cone as hypothesised by \cite{gopal16_typeIVdirectivity}. 
%\acknowledgement British  spelling: \verb+\acknowledgement+

%%%%%%%%%%%%%%%%%%%%%%%%%%%%%%%%%%%%%%%%%%%%%%%%%%%%%%%%%%%%%%%%%%%%%%%%%%%

\begin{acknowledgments}
We thank the CDAW team for maintaining an up to date catalog of LASCO/CME events and processed dynamic spectra from Wind and STEREO spacecraft. We also thank SWPC, Radio Monitoring, Australian Space Weather Data Center and eCallisto network for the open access database they maintain for various solar events. AM acknowledges the discussions with Sunpy team via Elements\footnote{\href{https://app.element.io/\#/room/\#sunpy:openastronomy.org}{https://app.element.io/\#/room/\#sunpy:openastronomy.org}} platform.  AK’s research was supported by an appointment to the NASA Postdoctoral Program at the the NASA Goddard Space Flight Center (GSFC).
AM and NG are supported in part by NASA's STEREO project and LWS program.
SA was partially supported by NSF grant, AGS-2043131.
SG's research was supported by NSF grant, AGS-22289.
AM acknowledges the developers of the various Python modules namely Numpy \citep{numpy}, Astropy \citep{astropy}, Matplotlib \citep{matplotlib} and multiprocessing. AM also thanks the developers of CASA \citep{casa}.
AM acknowledges the help from Seiji Yashiro in posting the catalog online and for maintaining the various CDAW data resources.
AM acknowledges Pertti M\"{a}kel\"{a} for useful  discussions and sharing of data.
\end{acknowledgments}

%% To help institutions obtain information on the effectiveness of their 
%% telescopes the AAS Journals has created a group of keywords for telescope 
%% facilities.
%
%% Following the acknowledgments section, use the following syntax and the
%% \facility{} or \facilities{} macros to list the keywords of facilities used 
%% in the research for the paper.  Each keyword is check against the master 
%% list during copy editing.  Individual instruments can be provided in 
%% parentheses, after the keyword, but they are not verified.

\vspace{5mm}
\facilities{Wind (WAVES), STEREO (SWAVES), SOHO (LASCO), SDO (AIA)}

%% Similar to \facility{}, there is the optional \software command to allow 
%% authors a place to specify which programs were used during the creation of 
%% the manuscript. Authors should list each code and include either a
%% citation or url to the code inside ()s when available.

\software{Numpy~\citep{numpy}, Astropy~\citep{astropy}, Matplotlib~\citep{matplotlib}, Multiprocessing~\citep{McKerns12_multiprocessing}, Pandas~\citep{reback20_pandas},
Urllib\footnote{\href{https://docs.python.org/3/library/urllib.html}{https://docs.python.org/3/library/urllib.html}}, Beautifulsoup\footnote{\href{https://www.crummy.com/software/BeautifulSoup/bs4/doc/}{https://www.crummy.com/software/BeautifulSoup/bs4/doc/}},
Sunpy~\citep{sunpy2020},
Datetime\footnote{\href{https://pypi.org/project/DateTime/}{https://pypi.org/project/DateTime/}},
Scipy~\citep{scipy}, Helioviewer~\citep{Heleioviewer2009}
}

%% Appendix material should be preceded with a single \appendix command.
%% There should be a \section command for each appendix. Mark appendix
%% subsections with the same markup you use in the main body of the paper.

%% Each Appendix (indicated with \section) will be lettered A, B, C, etc.
%% The equation counter will reset when it encounters the \appendix
%% command and will number appendix equations (A1), (A2), etc. The
%% Figure and Table counter will not reset.

\bibliography{allref}
\bibliographystyle{aasjournal}

%% This command is needed to show the entire author+affiliation list when
%% the collaboration and author truncation commands are used.  It has to
%% go at the end of the manuscript.
%\allauthors

%% Include this line if you are using the \added, \replaced, \deleted
%% commands to see a summary list of all changes at the end of the article.
%\listofchanges

\end{document}